\documentclass[aps, nofootinbib,twocolumn,superscriptaddress, preprintnumbers, longbibliography]{revtex4-1}
\pdfoutput=1

\usepackage{amsmath}
\usepackage{amsfonts}
\usepackage{amssymb}
\usepackage{bbm}
\usepackage{hyperref}
\usepackage{cleveref}
\usepackage{color}
\usepackage{comment}
\usepackage{caption}
\usepackage[latin9]{inputenc}
\usepackage{array}
\usepackage{booktabs}
\usepackage{breakurl}
\usepackage{subcaption}
\usepackage{ragged2e}

\usepackage{graphicx}
\usepackage{esint}



\begin{document}

\begin{flushright} 
\texttt{MPP-2018-303}
\end{flushright}

\title{Bimetric cosmology is compatible with local tests of gravity}

\author{Marvin L\"uben}
\email{mlueben@mpp.mpg.de}
\affiliation{Max-Planck-Institut f\"ur Physik (Werner-Heisenberg-Institut)\\
 F\"ohringer Ring 6, 80805 Munich, Germany}

\author{Edvard M\"ortsell}
\email{edvard@fysik.su.se}

\affiliation{Oskar Klein Centre, Stockholm University, AlbaNova University Center\\ 106 91 Stockholm, Sweden}
\affiliation{Department of Physics, Stockholm University, AlbaNova University Center\\ 106 91 Stockholm, Sweden}

\author{Angnis Schmidt-May}
\email{angnissm@mpp.mpg.de}

\affiliation{Max-Planck-Institut f\"ur Physik (Werner-Heisenberg-Institut)\\
 F\"ohringer Ring 6, 80805 Munich, Germany}

\begin{abstract}
Recently, Kenna-Allison \textit{et~al.}~claimed that bimetric gravity cannot give rise to a viable cosmological expansion history while at the same time being compatible with local gravity tests. In this note we review that claim and combine
various results from the literature to 
provide several simple counter examples. We conclude that the results of Kenna-Allison \textit{et~al.}~cannot hold in general. 
\end{abstract}

\maketitle

%%%%%%%%%%%%%%%%%%%%%%%%%%%%%%%%%%%
\section{Introduction}
%%%%%%%%%%%%%%%%%%%%%%%%%%%%%%%%%%%

Ghost-free bimetric gravity is an extension of general relativity (GR)
which describes the nonlinear interactions
of a massive spin-2 field in a dynamical gravitational 
background~\cite{Hassan:2011zd, Hassan:2012wr}.
In contrast to its cousin massive gravity~\cite{deRham:2010kj}, it contains
a massless spin-2 mode that mediates a long-range force and it possesses a 
smooth GR limit\footnote{An
important additional point to consider is the Cauchy problem, which is known
to be well-posed in GR, but still the subject of ongoing work within bimetric
theory~\cite{Kocic:2018ddp,Kocic:2018yvr,Kocic:2019zdy,Torsello:2019tgc}.}.

In the linear theory of a massive spin-2 field coupled to a source, the zero-mass limit is 
discontinuous~\cite{vanDam:1970vg, Zakharov:1970cc}. 
For a single massive graviton, this discontinuity is in conflict with
observations but can be cured by nonlinear interactions, a feature known as the
Vainshtein mechanism~\cite{Vainshtein:1972sx}. In massive gravity, without the 
massless spin-2 mode, this mechanism is absolutely crucial for the phenomenological viability of the
theory. In bimetric theory, the situation is fundamentally different since the 
gravitational interaction is mediated by a combination of the massive and the massless
spin-2 mode. 
For large spin-2 masses, bimetric gravity passes all local gravity tests
due to the strong Yukawa suppression of the massive graviton mode in the 
gravitational potential in Newtonian approximation. 
In addition, the coupling of the massive spin-2 mode to matter can be made arbitrarily
small, in which case the gravitational force is effectively mediated purely by the massless field.
As a consequence, bimetric theory does not necessarily require a working Vainshtein mechanism
in order to be in agreement with observational data.

Furthermore, bimetric theory can give rise to cosmological solutions with 
accelerated expansion even in the absence of vacuum energy~\cite{Comelli:2011zm, Volkov:2011an,vonStrauss:2011mq}. 
These self-accelerating solutions
require the mass scale which is associated to the breaking of independent diffeomorphisms of the two metrics
to be on the order of the Hubble scale. This scale is not necessarily identical to the Fierz-Pauli mass 
of the massive spin-2 mode since the latter depends on a different parameter combination. In particular,
the spin-2 mass also involves the parameter which controls the coupling of the massive spin-2 mode to the matter source.
 
The authors of Ref.~\cite{Kenna-Allison:2018izo} claimed that in bimetric theory it is not possible to bring a viable
cosmological expansion history in agreement with local gravity tests. Implicitly, their
argument involved two steps: (1) The assumption that local gravity tests can only be passed if the Vainshtein mechanism is 
at work, and (2) a working Vainshtein mechanism can be shown to contradict a viable background cosmology due to
incompatible requirements on the parameters of the theory.

In the following we invalidate both of these steps in the argument by giving explicit counter-examples.

%%%%%%%%%%%%%%%%%%%%%%%%%%%%%%%%%%%%%%%%%%%%%%%%%
\section{Short review of bimetric gravity}
%%%%%%%%%%%%%%%%%%%%%%%%%%%%%%%%%%%%%%%%%%%%%%%%%
We start by presenting the action of ghost-free bimetric theory for two metric tensors
$g$ and $f$ \cite{Hassan:2011zd,Hassan:2011ea},\footnote{For a review
on bimetric gravity, see Ref.~\cite{Schmidt-May:2015vnx}.} 
\begin{flalign}
	&S=\frac{m_g^2}{2}\int\text{d}^4 x\left(\sqrt{-g}\,R(g)+\alpha^2\sqrt{-f}\,R(f)\right)\nonumber\\
	&-m^2 m_g^2 \int\text{d}^4 x \,V(g,f) + \int\text{d}^4 x \sqrt{-g}\,\mathcal L_\text{m}(g,\Phi)\,,
\end{flalign}
where $m_g$ is the bare Planck mass of the metric $g$ and $\alpha$ 
measures the ratio to the bare Planck mass of $f$. 
$R(\bullet)$ denotes the Ricci curvature scalar for each metric. 
The two metric tensors interact via the potential,
\begin{flalign}
	V(g,f)=\sqrt{-g}\sum_{n=0}^4\beta_n e_n\left(\sqrt{g^{-1}f}\right)\,,
\end{flalign}
where $e_n$ denotes the $n$-th elementary symmetric polynomial of the matrix argument. 
The constants $\beta_n$ are free parameters of the theory, and $\beta_0$ and $\beta_4$ 
parametrize the bare cosmological constants for $g$ and $f$, 
respectively\footnote{Ref. \cite{Kenna-Allison:2018izo} refers to $\alpha_0$ 
as the bare cosmological constant, which is incorrect,
since matter loops renormalize $\beta_0$. For an explicit confirmation of this, see equation
(2.12) of Ref.~\cite{deRham:2014naa}. 
The parameter $\alpha_0$ is a combination of all five $\beta_n$ and
the exact relations between the two parametrizations can be found, for instance, in Ref.~\cite{Hassan:2012wr}.}.
%We work in the framework of \textit{singly-coupled} bimetric gravity, 
%where only the metric $g$ couples to matter fields, collectively denoted by $\Phi$.

Varying the action with respect to $g$ and $f$ yields two sets of modified Einstein equations,
\begin{subequations}
\begin{flalign}
	G^g_{\mu\nu}+m^2 V^g_{\mu\nu} =& \,\frac{1}{m_g^2}T_{\mu\nu}\,,\\
	\alpha^2 G^f_{\mu\nu} + m^2 V^f_{\mu\nu}=&\,0\,,
\end{flalign}
\end{subequations}
where $G^g$ and $G^f$ are the Einstein tensors of $g$ and $f$, respectively. The terms $V^{g,f}$ 
arising from variation of the potential were derived in Ref.~\cite{Hassan:2011vm, Hassan:2011zd}. 
Finally, the stress-energy tensor of matter is defined as,
\begin{flalign}
	T_{\mu\nu}=\frac{-2}{\sqrt{-g}}\frac{\delta\sqrt{-g}\,\mathcal L_\text{m}}{\delta g^{\mu\nu}}\,.
\end{flalign}

%%%%%%%%%%%%%%%%%%%%%%%%%%%%%%%%%%%%%%%%%%%%%%%%%%%%%%%%%%%%%%%%%%%%%%
\section{Mass eigenstates and gravitational force}
%%%%%%%%%%%%%%%%%%%%%%%%%%%%%%%%%%%%%%%%%%%%%%%%%%%%%%%%%%%%%%%%%%%%%%

Bimetric gravity has a well defined mass spectrum around proportional backgrounds, 
where the metrics are conformally related as $\bar f=c^2 \bar g$ with the real constant $c$,
determined by~\cite{Hassan:2012wr},
\begin{flalign}\label{bgeq}
{\alpha^2}\big(c\beta_0+3c^2\beta_1+3c^3\beta_2+c^4\beta_3\big)\nonumber\\
=\beta_1+3c\beta_2+3c^2\beta_3+c^3\beta_4\,.
\end{flalign}
We consider small fluctuations around that background,
\begin{flalign}
	g_{\mu\nu}=\bar g_{\mu\nu} + \delta g_{\mu\nu}\,, \qquad 
	f_{\mu\nu}=\bar f_{\mu\nu}+\delta f_{\mu\nu}\,.
\end{flalign}
Plugging this ansatz into the Einstein equations and 
keeping only terms up to linear order in the fluctuations, one finds that a massless spin-$2$ field, 
$\delta G_{\mu\nu}$, and a massive spin-$2$ field, $\delta M_{\mu\nu}$ propagate on the 
proportional background \cite{Hassan:2011zd,Hassan:2012wr}. 
The massive mode has a Fierz-Pauli mass,
\begin{flalign}\label{FPmass}
m_\text{FP}^2=m^2\frac{1+\alpha^2 c^2}{\alpha^2 c^2}c(\beta_1+2\beta_2 c+\beta_3 c^2)\,.
\end{flalign}
The original metric fluctuations are linear combinations of both mass eigenstates,
\begin{subequations}
\begin{flalign}
	&\delta g_{\mu\nu}\propto \delta G_{\mu\nu} - \alpha^2 \delta M_{\mu\nu}\,,
	\label{massmode}\\
	&\delta f_{\mu\nu}\propto \delta M_{\mu\nu} + c^{2} \delta G_{\mu\nu}\,,
\end{flalign}
\end{subequations}
where we omitted the overall normalization.

When $\alpha\ll1$, the fluctuation of the physical metric $g_{\mu\nu}$ 
is almost aligned with the massless excitation. Since matter fields couple to the metric perturbations 
$\delta g_{\mu\nu}$, we expect to recover GR for $\alpha\ll1$ \cite{Akrami:2015qga}. 
Clearly, in this parameter region there is no conflict with current observational data.

For instance, let us consider a spherically symmetric 
background \cite{Babichev:2013pfa,Comelli:2011wq,Enander:2015kda,Enander:2013kza,Babichev:2016bxi}.
 The contribution to the Newtonian potential coming from the
 massless mode is a Coulomb-like term, $\sim r^{-1}$, proportional to the inverse distance
 between the source and the test particle, 
 while the massive mode contributes a Yukawa-term $\sim e^{-m_\text{FP}r}/r$. Hence,
 from eq.~(\ref{massmode}) it follows that
 the coupling $\frac{1}{m_g^2}\delta g_{\mu\nu} T^{\mu\nu}$ will produce the following
 linearized gravitational potential,
\begin{flalign}\label{modpot}
	V(r)=-\frac{1}{m_\text{Pl}^2}\left(\frac{1}{r}+\frac{4\alpha^2c^2}{3}\frac{e^{-m_\text{FP}r}}{r}\right),
\end{flalign}
where the physical Planck mass is $m_\text{Pl}^2=m_g^2(1+\alpha^2 c^2)$~\cite{Hassan:2012wr}. 
Whenever the first term in the gravitational potential dominates over the second one,
the solution behaves approximately like GR and does not require a Vainshtein mechanism.
We can thus identify two parameter 
r\'egimes in which GR is restored,
\begin{itemize}
	\item[1.)] $\alpha c\ll1$,
	\item[2.)] $m_\text{FP}\gg \ell^{-1}$,
\end{itemize}
where $\ell$ is the typical length scale of the system (e.g. $\ell\simeq$\,1\,AU for the Solar System).

%%%%%%%%%%%%%%%%%%%%%%%%%%%%%%%%%%%%%%%%%%%%%%%%%%%%%%%
\section{Example: Solar System}
%%%%%%%%%%%%%%%%%%%%%%%%%%%%%%%%%%%%%%%%%%%%%%%%%%%%%%%

In this section, we demonstrate the general results
discussed in the previous section in an explicit example: the Solar System. 
For concreteness, we derive numerical values for the Sun as central source of the
gravitational potential. Not all Solar System tests are based on this scenario,
so our arguments here should be viewed as qualitative. A detailed quantitative analysis
is left for future work.
Our findings are summarised in \cref{fig:inclusion-plot} and \cref{fig:alpha-mass-inclusion-plot}.

\subsection{Yukawa suppression}

\begin{figure}
	\centering
	\includegraphics[scale=0.66]{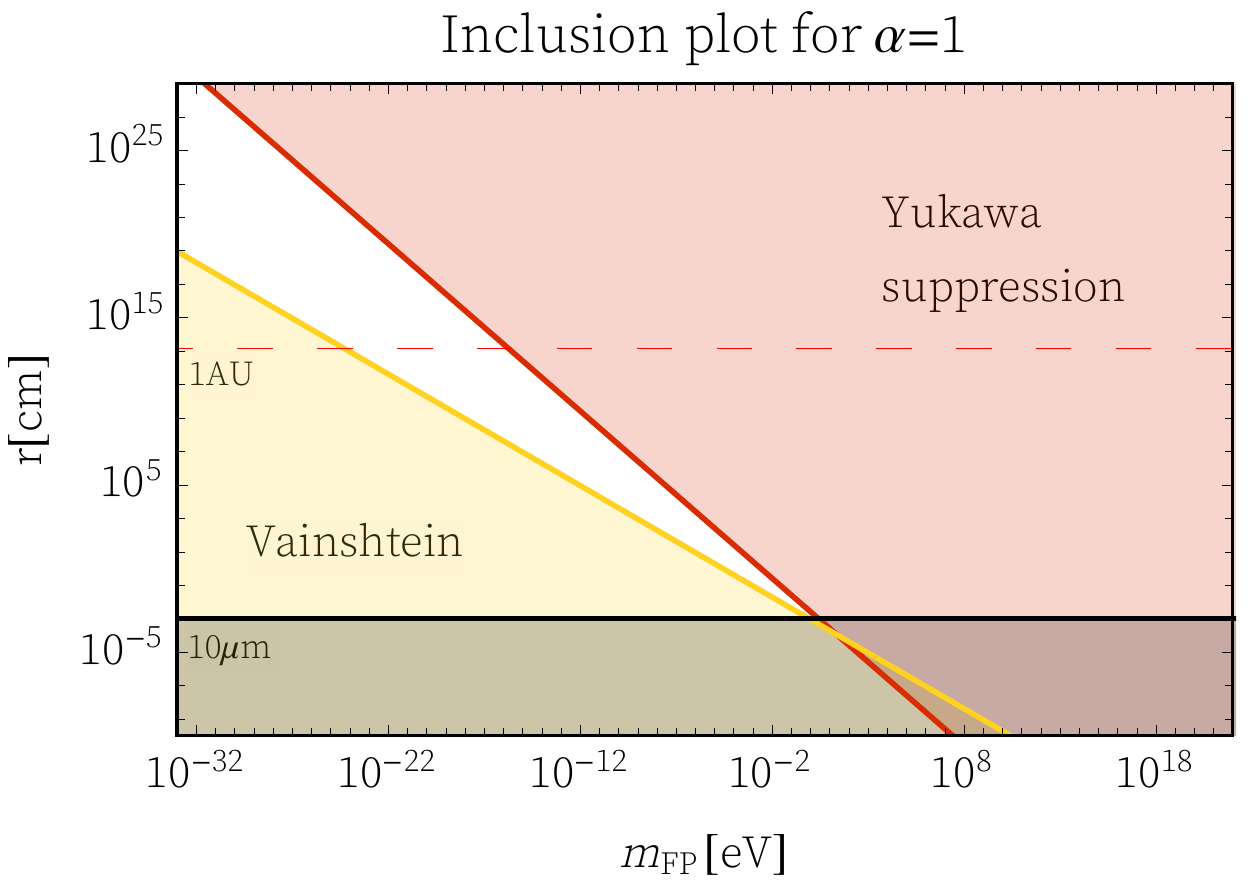}
	\caption{\raggedright For the Sun as central source, this figure indicates deviations from GR in the Newtonian force, for distances $r$ to the test particle as a function of the spin-2 mass $m_\text{FP}$. For large $m_\text{FP}$ and large $r$, the contribution from the massive mode is sufficiently suppressed, yielding the red-shaded region. For small $m_\text{FP}$ and small $r$, the Vainshtein mechanism restores GR, yielding the yellow-shaded region. Below $10\,\mu\text{m}$ no observational constraints exist. }
	\label{fig:inclusion-plot}
\end{figure}

At the scale of $1\,\text{AU}\simeq 1.5\times 10^{13} \,\text{cm}$, deviations from the 
inverse square law for the gravitational force are constrained to 
be $\lesssim 10^{-9}$ \cite{Will:2014kxa}. We aim at providing the most conservative
 constraints on the bimetric parameters. Hence, we use this bound (which is the most stringent one) 
 on deviations for any distance from the Sun. Comparing the two contributions in \eqref{modpot}, 
 this requires\footnote{In this section, we absorb $c$ into $\alpha$ for simplicity. Reinserting $c$ does not change any of our findings.},
\begin{flalign}
	\frac{4\alpha^2}{3} e^{-m_\text{FP}r}\lesssim 10^{-9}.\label{eq:SS-bound}
\end{flalign}
Deviations due to the massive mode are arbitrarily small in both the previously 
identified parameter limits. The red line in \cref{fig:inclusion-plot} represents the 
bound \cref{eq:SS-bound} for the case $\alpha=1$. Therefore, in the red-shaded region, 
the Yukawa-like term in the gravitational potential is always smaller than all observational bounds.
A value for $\alpha$ smaller than unity makes the deviations even smaller.
 
Local gravity tests inside the Solar System provide strong 
constraints on deviations from GR 
down to scales of $\sim10\,\mu\text{m}$ \cite{Will:2014kxa}. 
As a very rough estimate, we use \cref{eq:SS-bound} to define a critical spin-2 mass (for $\alpha\simeq 1$),
\begin{flalign}
	m_{\text{crit}}\simeq 2.6\,\text{eV},
\end{flalign}
above which no deviations from GR are detectable via observations in the Solar System.
This is indicated by the black-shaded region in \cref{fig:inclusion-plot}. 

For $m_{\text{FP}}\gg m_{\text{crit}}$ 
and in the red-shaded region of effective Yukawa suppression, 
the Vainshtein mechanism is certainly not needed to restore GR.

\subsection{Vainshtein regime}

Close to the source, nonlinear terms become as important as 
the linear ones in a way, such that GR is restored. 
This is the well-known Vainshtein mechanism \cite{Vainshtein:1972sx} 
and it is active within a sphere defined by the Vainshtein radius,
\begin{flalign}\label{VSrad}
	r_V=\left(\frac{r_S}{m_\text{FP}^2}\right)^{1/3}
\end{flalign}
around an object of mass $M$ with Schwarzschild radius 
$r_S=(1+\alpha^2) M/m_\text{Pl}^2$ \cite{Babichev:2013pfa,Babichev:2016bxi}.
Even though the expression derived in Ref.~\cite{Kenna-Allison:2018izo} 
seems to differ by a numerical factor (which is not manifestly positive) from \eqref{VSrad}, the scale is the same
as long as the parameters satisfy $\beta_n\sim\mathcal{O}(1)$.\footnote{We emphasize that the 
value for the Vainshtein radius is derived assuming $r<m_\text{FP}^{-1}$. For the Solar System,
this is not an issue since the intersection of the lines corresponding to $r_V$ and $m_\text{FP}^{-1}$ lies close to the observable threshold of $\sim 10\,\mu\text{m}$. The region where $m_\text{FP}^{-1}\sim r_V$ should be treated more carefully, but can definitely be brought in agreement with data by making $\alpha$ small.}

Well inside the Vainshtein radius, deviations from the inverse-square law of the gravitational force scale like 
$(r/r_V)^3$~\cite{Babichev:2013pfa,Enander:2015kda} and hence, in order to satisfy
\eqref{eq:SS-bound}, we require (for $\alpha\simeq 1$),
\begin{flalign}
	\frac{r}{r_V}\lesssim 10^{-3}\,.\label{eq:Vainshtein-bound}
\end{flalign}
Choosing a solar mass object $M=M_\odot$ defines the orange line in \cref{fig:inclusion-plot} as a rough estimate for the 
threshold of an effective Vainshtein mechanism. Consequently, well inside the yellow-shaded region,
 the gravitational force is certainly indistinguishable from GR 
 for any bimetric parameters.

\begin{figure}
	\centering
	\includegraphics[scale=0.66]{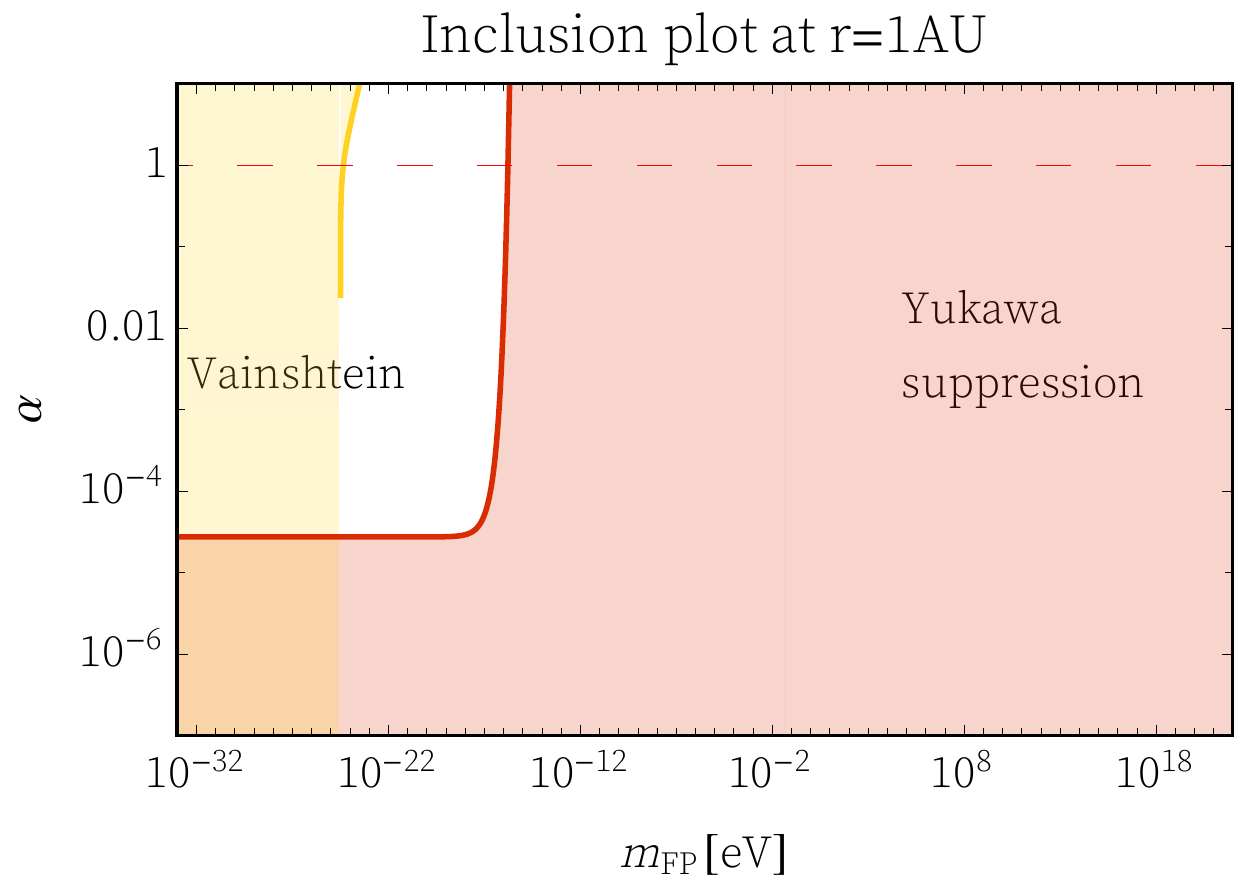}
	\caption{ \raggedright This figure shows the allowed parameter regions in the $\alpha$-$m_\text{FP}$-plane at the length scale $r=1\text{AU}$. For large $m_\text{FP}$, the parameter $\alpha$ is not constrained to be small thanks to the Yukawa suppression in the potential. For small $m_\text{FP}$, it is unconstrained due to the Vainshtein mechanism. Only for masses from $10^{-24}$ to $10^{-16}$\,eV observations require $\alpha\lesssim 10^{-5}$.}
	\label{fig:alpha-mass-inclusion-plot}
\end{figure}

\subsection{Constraints on the spin-2 coupling}

Only in the region left white in \cref{fig:inclusion-plot}, 
significant deviations from GR could occur. However, we still have the free parameter 
$\alpha$, which we can use to suppress the extra term in the Newtonian potential. 
The most stringent bound comes from close to the Vainshtein radius, where the 
observational bound~\eqref{eq:SS-bound} requires,
\begin{flalign}
	\alpha\lesssim10^{-5},\label{eq:alpha-bound}
\end{flalign}
see also \cref{fig:alpha-mass-inclusion-plot}.
Then, for any mass of the spin-2, deviations from the GR prediction are 
undetectable with current experimental precision.

For very small spin-2 masses, the Vainshtein radius of the Sun becomes larger than the 
Solar System itself. As a conservative estimate, evaluating \cref{eq:Vainshtein-bound} 
at $r=100\,\text{AU}$ implies $m_\text{FP}\lesssim 5\times 10^{-28}\text{eV}$.
 For smaller spin-2 masses the Kuiper belt is well inside the Vainshtein sphere and 
 the Solar System does not constrain $\alpha$ to be small.

We conclude this section by stressing the following points:
For a large spin-2 mass, $m_\text{FP}\gg m_\text{crit}$, 
the Vainshtein mechanism is not necessary for the theory to pass all Solar System tests
since the Yukawa suppression of the contribution from the massive mode is too large. 
For very small masses, the Vainshtein mechanism has to be (and is) active.
These results hold for any value of $\alpha\lesssim 1$.
Only for intermediate spin-2 masses the parameter $\alpha$ has to satisfy the bound~\eqref{eq:alpha-bound}.

%%%%%%%%%%%%%%%%%%%%%%%%%%%%%%%%%%%%%%%%%%%%%%%%%%%%%%%%%%%%%%%%%%%%%%
\section{Gravitational waves and galactic tests}\label{sec:other-constraints}
%%%%%%%%%%%%%%%%%%%%%%%%%%%%%%%%%%%%%%%%%%%%%%%%%%%%%%%%%%%%%%%%%%%%%%
Constraints from gravitational waves were derived in Ref.~\cite{Max:2017flc}, excluding a small region of the parameter space: Within the spin-$2$ mass range of $10^{-22}$ to $10^{-21}\text{eV}$, the ratio of the Planck masses has to satisfy $\alpha\lesssim0.3$. For a spin-$2$ mass outside that range, gravitational wave observations do not constrain $\alpha$ to be small. 

On galactic scales, observations of galactic velocity dispersions and gravitational lensing can be used to constrain 
the bimetric parameters \cite{Enander:2013kza}. For $m_\text{FP}\simeq H_0$ and for a galaxy of mass 
$M=10^{12} M_\odot$, deviations from GR predictions are expected to be on the order of
 $(r/r_\mathrm{V})^3\simeq 2\times 10^{-13}$~\cite{Enander:2015kda}, which is within the observational 
bounds by a factor of $4\times 10^{-12}$~\cite{Schwab:2009nz,Sjors:2011iv}.
Data from cluster lensing requires $\alpha\lesssim0.7$ for a spin-$2$ mass range of 
$10^{-30}$ to $10^{-28}\text{eV}$~\cite{Platscher:2018voh}, where neither the Vainshtein mechanism
nor the Yukawa suppression restore GR on the length scales relevant to observations.
Outside of this range, 
$\alpha$ is not constrained to be small since it is not needed to restore GR.

For a combined exclusion plot with galactic and extragalactic constraints as 
well as bounds from gravitational wave detection on the bimetric parameters, 
see fig.~9 in Ref.~\cite{Platscher:2018voh}.\footnote{Note that $\alpha=\tan\theta$ 
in the notation of Ref.~\cite{Platscher:2018voh} 
and their plot is linear in $\theta$.}
We conclude that tests of the gravitational law outside the Solar System 
do not provide constraints on the bimetric parameter space in the region where $\alpha\lesssim0.1$.
Thus the most stringent bounds on bimetric gravity
are those from the Solar System, discussed in the previous section.

%%%%%%%%%%%%%%%%%%%%%%%%%%%%%%%%%%%%%%%%%%%%%%%%%%%%%%%%%%%%%%%%%%%%%%
\section{Bimetric cosmologies}
%%%%%%%%%%%%%%%%%%%%%%%%%%%%%%%%%%%%%%%%%%%%%%%%%%%%%%%%%%%%%%%%%%%%%%

The equations of motion of bimetric gravity evaluated on a homogeneous and
isotropic ansatz for both metrics give two sets of equations that can be combined
into one modified Friedmann equation of the general form,
\begin{equation}\label{modfr}
H^2=F(\rho)\,.
\end{equation}
Here $H=\dot{a}/a$ is the Hubble function of the physical scale factor $a(t)$ 
inside the physical metric $g_{\mu\nu}$ and $\rho$ is the matter energy density, i.e.,~the 00-component of the 
perfect fluid source $T^\mu_{~\nu}$.

In the following, we will assume that all $\beta_n$ parameters,
are of order unity and the mass scale $m$ (not necessarily $m_\mathrm{FP}$) is of order $H_0$.
For concreteness we will consider a simple bimetric model with
parameters $\beta_0=\beta_3=\beta_4=0$. 
All of our findings straightforwardly generalize to models with generic $\mathcal{O}(1)$ values for all $\beta_n$.

\subsection{Large mass region}
In Ref.~\cite{Akrami:2015qga} it was shown that for $\alpha\ll 1$, the Friedmann equation~(\ref{modfr})
can be approximated by the simple form,
\begin{equation}\label{Hexp}
H^2=\frac{\rho}{3m_\text{Pl}^2}-\frac{2\beta_1^2}{3\beta_2}m^2+\mathcal{O}(\alpha^2)\,.
\end{equation}
The Fierz-Pauli mass in eq.~\eqref{FPmass}, 
defined at the de Sitter point is,
\begin{equation}
m_\mathrm{FP}^2= m^2(1+\alpha^{-2} c^{-2})(c\beta_1+2c^2\beta_2)\sim \frac{m^2}{\alpha^2}\,,
\end{equation}
where we have used that $c=-\frac{\beta_1}{3\beta_2}+\mathcal{O}(\alpha^2)\sim\mathcal{O}(1)$ 
which follows from the background equation~\eqref{bgeq}.
Clearly $\alpha\ll 1$ implies that,
\begin{equation}
m_\mathrm{FP}\gg H_0\,.
\end{equation}
This shows that we can easily be in the parameter regime where no Vainshtein mechanism is needed and 
at the same time have a valid background cosmology that matches exactly that of GR with
cosmological constant $\Lambda=2m^2\beta_1^2/{\left|\beta_2\right|}$.
 In fact, the 
approximation in eq.~(\ref{Hexp}) is valid up to the scale $H\sim m_\mathrm{FP}$~\cite{Akrami:2015qga}.
By making the spin-2 mass large, we can thus push back the deviations from GR to arbitrary early times.
Explicitly, already for a spin-$2$ mass of $m_\text{FP}\simeq 0.4\,\text{eV}$ the background expansion in this simple 
model follows (almost) exactly the $\Lambda$CDM prediction until the CMB.
At earlier times, when $H> m_\mathrm{FP}$, we expect GR to be recovered by
nonlinear effects, as we shall discuss further below. Moreover, the cosmological evolution
at early times is dominated by the matter density $\rho$~\cite{vonStrauss:2011mq} and will therefore
anyway follow almost exactly the evolution predicted by the $\Lambda$CDM model.

\subsection{Small mass region}
Now, we consider the parameter region, 
where $\alpha\simeq 1$. With our assumption $\beta_n\sim\mathcal O(1)$, this implies
\begin{flalign}
	m_\text{FP}\simeq H_0.
\end{flalign}
This parameter region has received a lot of attention in the 
literature, see e.g.~Ref.~\cite{Comelli:2011zm, Volkov:2011an,vonStrauss:2011mq,Akrami:2012vf, Konnig:2013gxa}.
In Ref.~\cite{Akrami:2012vf} various models with $\alpha=1$ were compared to observations, 
many of which give a good fit to data while still allowing for testable deviations from the standard $\Lambda$CDM
scenario. 
The crucial point is that this parameter region is not 
in conflict with a successful Vainshtein mechanism around spherically symmetric sources. 
For a spin-2 mass on the order of the Hubble rate today, the Vainshtein radius of the sun 
is $r_V\simeq 10^{22}\,\text{cm}$, which is on the order of the size of the Milky Way. 
Hence, no constraints on $\alpha$ exist for such a tiny spin-2 mass.
Lensing constraints do not provide bounds on the value of $\alpha$ 
for small spin-2 masses either \cite{Platscher:2018voh}, see \cref{sec:other-constraints}.
The only caveat here is that the effective Planck mass in cosmological solutions must be matched with
Newton's constant in the gravitational force law, which may (mildly) constrain one of the
$\beta_n$ parameters.

\subsection{Cosmological perturbations}
Although bimetric gravity gives rise to a viable expansion history while passing 
all local gravity tests in both these parameter regions, the cosmological perturbations behave differently
from GR. 
In particular, the FLRW background is not 
always stable against scalar fluctuations during the entire expansion history due to
a gradient instability~\cite{Comelli:2012db,Khosravi:2012rk,Berg:2012kn,Fasiello:2013woa,Comelli:2014bqa,DeFelice:2014nja,Solomon:2014dua,Lagos:2014lca,Konnig:2014dna,Konnig:2014xva,Enander:2015vja,Konnig:2015lfa}. 
Ref.~\cite{Akrami:2015qga} showed that for times earlier than,
\begin{flalign}
	H\simeq m_\text{FP}\,,
\end{flalign}
the instability sets in. 
Thus, by making the spin-2 mass large (which is equivalent to $\alpha\ll1$ in our parametrization),
the instabilities can be pushed to arbitrarily early times.

Na\"ively, these results seem to disfavour the parameter region with small spin-2 mass \cite{Konnig:2014dna}.
However, precisely when the Hubble rate exaggerates the spin-2 mass, 
the inset of the gradient instability simply implies that nonlinear effects 
become as important as linear ones, invalidating linear perturbation theory. 
The results of Ref.~\cite{Mortsell:2015exa, Aoki:2015xqa} suggest that these nonlinear effects
are in fact not problematic but instead restore GR at early times through the Vainshtein mechanism.

%%%%%%%%%%%%%%%%%%%%%%%%%%%%%%
\section{Conclusion}
%%%%%%%%%%%%%%%%%%%%%%%%%%%%%%
We have discussed on general grounds why bimetric theory for a large range of parameter values can 
give rise to a viable cosmology while at the same time passing all other tests of the gravitational force law. 

In particular, we demonstrated explicitly that for a weak coupling of the massive spin-2 mode to matter
and a large spin-2 mass, bimetric theory becomes essentially indistinguishable from GR in cosmological 
solutions and on all distance scales in the galaxy. 

We emphasize once more that this is in sharp contrast to massive gravity, which can be obtained as a 
parameter limit in bimetric gravity by sending $\alpha\rightarrow \infty$ and which requires
the Vainshtein mechanism. For a summary of the much stronger constraints
on massive gravity, see Ref.~\cite{deRham:2016nuf}.

We also showed that even in the different regime 
of small spin-2 mass and Planck scale coupling for the massive spin-2, bimetric theory is compatible
with all available observational data. 

The assumptions in our explicit examples may seem restrictive. 
%Including more interaction parameters allows to move to other, more extreme regions of the 
%parameter space that require tuning among the $\beta_n$. 
The theory has enough free 
parameters to achieve a large spin-2 mass even when $\alpha\simeq1$ and conversely, 
a small spin-2 mass $m_\text{FP}\simeq H_0$ although $\alpha\ll1$. These more general 
models, which require values for the $\beta_n$ parameters different from $\mathcal{O}(1)$,  
have several branches of solutions and the expressions can be lengthy. 
It would be interesting to study the resulting phenomenology, which we leave for 
future work. It is possible that the results 
of Ref.~\cite{Kenna-Allison:2018izo} apply to these parameter regions. 
Nevertheless, the examples we presented here clearly show that they
do not hold in general.

We conclude that bimetric cosmologies are certainly not incompatible with local gravity tests.

\vspace{5pt}
\subsubsection*{Acknowledgments} This work is supported by a grant from the Max-Planck-Society.

\bibliography{CosmoFinal}

\end{document}